\documentclass[superscriptaddress,prc,twocolumn,preprintnumbers,amsmath,amssymb,bibnotes,altaffilletter,floatfix,nofootinbib]{revtex4-2}

\usepackage{graphicx}
\usepackage{dcolumn}
\usepackage{bm}
\DeclareUnicodeCharacter{2212}{-}

\usepackage{lineno}

\usepackage[colorlinks,citecolor=blue,linkcolor=blue,urlcolor=blue]{hyperref}%

\usepackage{natbib}
\usepackage{mfirstuc}
\usepackage{makecell}

\newcommand{\demo}{\textsc{Demonstrator}}
\newcommand{\mjd}{\textsc{Majorana Demonstrator}}
\newcommand{\mc}{\textsc{Majorana} Collaboration}

\begin{document}


\title{Experimental study of $^{13}$C($\alpha, n)^{16}$O reactions in the {\sc Majorana Demonstrator} calibration data}

\newcommand{\ITEP}{National Research Center ``Kurchatov Institute'' Institute for Theoretical and Experimental Physics, Moscow, 117218 Russia}
\newcommand{\JINR}{Joint Institute for Nuclear Research, Dubna, 141980 Russia} 
\newcommand{\lbnl}{Nuclear Science Division, Lawrence Berkeley National Laboratory, Berkeley, CA 94720, USA}
\newcommand{\lbnle}{Engineering Division, Lawrence Berkeley National Laboratory, Berkeley, CA 94720, USA}
\newcommand{\lanl}{Los Alamos National Laboratory, Los Alamos, NM 87545, USA}
\newcommand{\queens}{Department of Physics, Engineering Physics and Astronomy, Queen's University, Kingston, ON K7L 3N6, Canada}
\newcommand{\uw}{Center for Experimental Nuclear Physics and Astrophysics, and Department of Physics, University of Washington, Seattle, WA 98195, USA}
\newcommand{\unc}{Department of Physics and Astronomy, University of North Carolina, Chapel Hill, NC 27514, USA}
\newcommand{\duke}{Department of Physics, Duke University, Durham, NC 27708, USA}
\newcommand{\ncsu}{Department of Physics, North Carolina State University, Raleigh, NC 27695, USA}	
\newcommand{\ornl}{Oak Ridge National Laboratory, Oak Ridge, TN 37830, USA}
\newcommand{\ou}{Research Center for Nuclear Physics, Osaka University, Ibaraki, Osaka 567-0047, Japan}
\newcommand{\pnnl}{Pacific Northwest National Laboratory, Richland, WA 99354, USA}
\newcommand{\ttu}{Tennessee Tech University, Cookeville, TN 38505, USA}
\newcommand{\sdsmt}{South Dakota School of Mines and Technology, Rapid City, SD 57701, USA}
\newcommand{\usc}{Department of Physics and Astronomy, University of South Carolina, Columbia, SC 29208, USA}
\newcommand{\usd}{Department of Physics, University of South Dakota, Vermillion, SD 57069, USA}  
\newcommand{\ut}{Department of Physics and Astronomy, University of Tennessee, Knoxville, TN 37916, USA}
\newcommand{\tunl}{Triangle Universities Nuclear Laboratory, Durham, NC 27708, USA}
\newcommand{\mpi}{Max-Planck-Institut f\"{u}r Physik, M\"{u}nchen, 80805, Germany}
\newcommand{\tum}{Physik Department and Excellence Cluster Universe, Technische Universit\"{a}t, M\"{u}nchen, 85748 Germany}
\newcommand{\williams}{Physics Department, Williams College, Williamstown, MA 01267, USA}
\newcommand{\ciemat}{Centro de Investigaciones Energ\'{e}ticas, Medioambientales y Tecnol\'{o}gicas, CIEMAT 28040, Madrid, Spain}
\newcommand{\iu}{Department of Physics, Indiana University, Bloomington, IN 47405, USA}
\newcommand{\iuceem}{IU Center for Exploration of Energy and Matter, Bloomington, IN 47408, USA}

\author{I.J.~Arnquist}\affiliation{\pnnl} 
\author{F.T.~Avignone~III}\affiliation{\usc}\affiliation{\ornl}
\author{A.S.~Barabash}\affiliation{\ITEP}
\author{C.J.~Barton}\affiliation{\usd}	
\author{K.H.~Bhimani}\affiliation{\unc}\affiliation{\tunl}
\author{E.~Blalock}\affiliation{\ncsu}\affiliation{\tunl}{} 
\author{B.~Bos}\affiliation{\unc}\affiliation{\tunl} 
\author{M.~Busch}\affiliation{\duke}\affiliation{\tunl}	
\author{M.~Buuck}\altaffiliation{SLAC National Accelerator Laboratory, Menlo Park, CA 94025, USA} \affiliation{\uw}
\author{T.S.~Caldwell}\affiliation{\unc}\affiliation{\tunl}	
\author{Y-D.~Chan}\affiliation{\lbnl}
\author{C.D.~Christofferson}\affiliation{\sdsmt} 
\author{P.-H.~Chu}\affiliation{\lanl} 
\author{M.L.~Clark}\affiliation{\unc}\affiliation{\tunl} 
\author{C.~Cuesta}\affiliation{\ciemat}	
\author{J.A.~Detwiler}\affiliation{\uw}	
\author{Yu.~Efremenko}\affiliation{\ut}\affiliation{\ornl}
\author{H.~Ejiri}\affiliation{\ou}
\author{S.R.~Elliott}\affiliation{\lanl}
\author{G.K.~Giovanetti}\affiliation{\williams}  
\author{M.P.~Green}\affiliation{\ncsu}\affiliation{\tunl}\affiliation{\ornl}   
\author{J.~Gruszko}\affiliation{\unc}\affiliation{\tunl} 
\author{I.S.~Guinn}\affiliation{\unc}\affiliation{\tunl} 
\author{V.E.~Guiseppe}\affiliation{\ornl}	
\author{C.R.~Haufe}\affiliation{\unc}\affiliation{\tunl}	
\author{R.~Henning}\affiliation{\unc}\affiliation{\tunl}
\author{D.~Hervas~Aguilar}\affiliation{\unc}\affiliation{\tunl} 
\author{E.W.~Hoppe}\affiliation{\pnnl}
\author{A.~Hostiuc}\affiliation{\uw} 
\author{M.F.~Kidd}\affiliation{\ttu}	
\author{I.~Kim}\affiliation{\lanl} 
\author{R.T.~Kouzes}\affiliation{\pnnl}
\author{T.E.~Lannen~V}\affiliation{\usc} 
\author{A.~Li}\affiliation{\unc}\affiliation{\tunl}
\author{A.M.~Lopez}\affiliation{\ut}	
\author{J.M. L\'opez-Casta\~no}\affiliation{\ornl} 
\author{E.L.~Martin}\altaffiliation{Present address: Duke University, Durham, NC 27708}\affiliation{\unc}\affiliation{\tunl}	
\author{R.D.~Martin}\affiliation{\queens}	
\author{R.~Massarczyk}\affiliation{\lanl}		
\author{S.J.~Meijer}\affiliation{\lanl}	
\author{T.K.~Oli}~\email{tupendra.oli@coyotes.usd.edu}\affiliation{\usd}  
\author{G.~Othman}\altaffiliation{Present address: Universit{\"a}t
Hamburg, Institut f{\"u}r Experimentalphysik, Hamburg,
Germany}\affiliation{\unc}\affiliation{\tunl}
\author{L.S.~Paudel}\affiliation{\usd} 
\author{W.~Pettus}\affiliation{\iu}\affiliation{\iuceem}	
\author{A.W.P.~Poon}\affiliation{\lbnl}
\author{D.C.~Radford}\affiliation{\ornl}
\author{A.L.~Reine}\affiliation{\unc}\affiliation{\tunl}	
\author{K.~Rielage}\affiliation{\lanl}
\author{N.W.~Ruof}\affiliation{\uw}	
\author{D.~Tedeschi}\affiliation{\usc}		
\author{R.L.~Varner}\affiliation{\ornl}  
\author{S.~Vasilyev}\affiliation{\JINR}	
\author{J.F.~Wilkerson}\affiliation{\unc}\affiliation{\tunl}\affiliation{\ornl}    
\author{C.~Wiseman}\affiliation{\uw}		
\author{W.~Xu}\affiliation{\usd} 
\author{C.-H.~Yu}\affiliation{\ornl}
\author{B.X.~Zhu}\altaffiliation{Present address: Jet Propulsion Laboratory, California Institute of Technology, Pasadena, CA 91109, USA}\affiliation{\lanl}

\collaboration{{\sc{Majorana}} Collaboration}
\noaffiliation

\date{\today} 


\begin{abstract}
Neutron captures and delayed decays of reaction products are common sources of backgrounds in ultra-rare event searches. In this work, we studied $^{13}$C($\alpha,n)^{16}$O reactions induced by $\alpha$-particles emitted within the calibration sources of the \mjd. These sources are thorium-based calibration standards enclosed in carbon-rich materials. The reaction rate was estimated by using the 6129-keV $\gamma$-rays emitted from the excited $^{16}$O states that are populated when the incoming $\alpha$-particles exceed the reaction Q-value. Thanks to the excellent energy performance of the \textsc{Demonstrator}'s germanium detectors, these characteristic photons can be clearly observed in the calibration data. Facilitated by \textsc{Geant4} simulations, a comparison between the observed 6129-keV photon rates and predictions by a TALYS-based software was performed. The measurements and predictions were found to be consistent, albeit with large statistical uncertainties. This agreement provides support for background projections from ($\alpha,n$)-reactions in future double-beta decay search efforts.
\end{abstract}


\maketitle 

\section{\label{sec:level1}Introduction}

Neutron-related reactions are an important source of background in underground neutrino and dark matter experiments~\cite{carson2004neutron,cooley2018input, chen2021radiogenic, Febbraro2020}. One common source of neutrons is ($\alpha,n$) reactions. Neutrons may penetrate shielding layers before being captured on sensitive detector materials, often creating radioactive isotopes, the delayed decays of which could be difficult to reject due to a lack of coincidence timing information. For example, in germanium-based neutrinoless double-beta decay ($0\nu\beta\beta$) experiments, neutron captures on $^{76}$Ge create $^{77}$Ge (half-life: 11.3 hr) and $^{77m}$Ge (half-life: 53.7 s) isotopes. The $\beta$-decay of these isotopes could potentially produce signals similar to $0\nu\beta\beta$ and with energies near the double-beta decay Q-value (Q$_{\beta\beta}$) of $^{76}$Ge. This background has been studied in detail~\cite{wiesinger2018virtual,arnquist2022signatures}. 

$^{232}$Th and $^{238}$U decay chains contain several $\alpha$-emitters. These naturally-occurring isotopes are present in detector materials, and various $\alpha$-particles with energies up to 9 MeV are emitted, initiating a range of $(\alpha,n)$ reactions. Even though the cleanest materials can be assayed and selected \cite{abgrall2016majorana}, stringent background requirements, especially for future experiments, demand an understanding of these neutron contributions with reasonable detail and precision. In particular, different types of plastic materials are widely used in low-background experiments, \textit{e.g.} for electrical insulation and neutron shielding. In these carbon-rich plastic materials, the $^{13}$C($\alpha, n)^{16}$O reaction is a major source of neutrons. 
  
  Besides its role as a background, the $^{13}$C($\alpha, n)^{16}$O reaction is considered the most important neutron source for s-process nucleosynthesis in low-mass asymptotic giant branch stars~\cite{busso2001nucleosynthesis,guo2012new,la2013measurement,Cristallo2018,Arnould2020}. This reaction and its cross section have been studied, and the results agree reasonably well among different measurements for low-energy $\alpha$-particles below about 5~MeV~\cite{sekharan1967c,davids1968study,heil200813,bair1973total,drotleff1993reaction,harissopulos2005cross, Broggini2018}. At higher $\alpha$ energies, precise cross section measurements are sparse~\cite{harissopulos2005cross,peters2017comment}, although new studies have been published for the 5-to-6~MeV region ~\cite{Febbraro2020} and more measurements are planned in the near future~\cite{Broggini2018}.   
 
\begin{figure}[htbp]
\includegraphics[width=\linewidth]{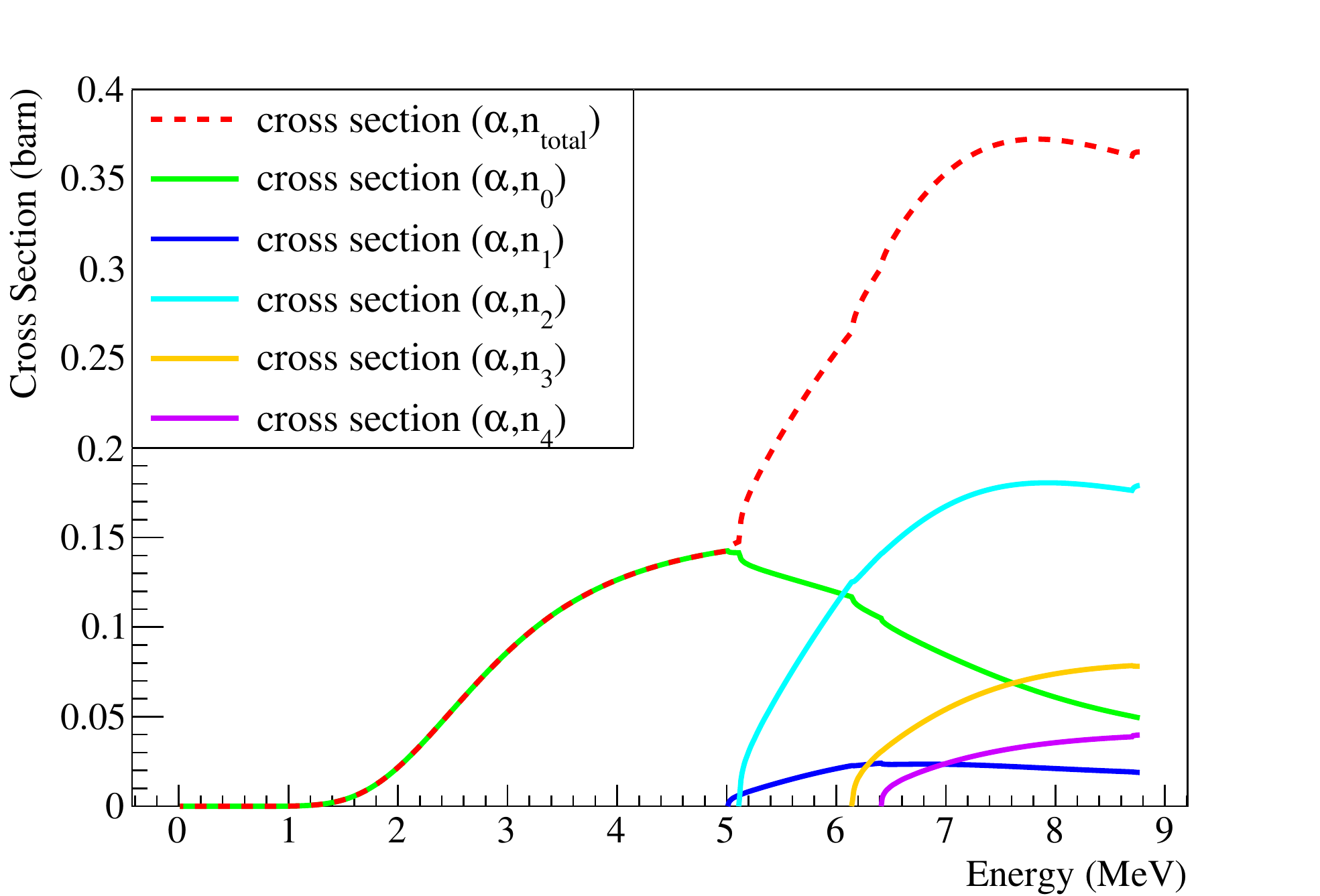}
\caption{Total cross section and the partial cross sections for the $^{13}$C($\alpha,n)^{16}$O reactions as a function of incident $\alpha$-particle energy available from the decay chain of $^{228}$Th. These cross sections are generated by TALYS-1.95. The results of the new TALYS version are consistent with branching ratios obtained from Ref.~\cite{mohr2018revised} that used  TALYS-1.8.}
\label{fig:cross_13C}
\end{figure}

In addition to relying on measured data, one can obtain ($\alpha,n$) cross sections from a statistical modeling approach using a nuclear reaction code such as TALYS~\cite{koning2013talys}. The TALYS-generated Evaluated Nuclear Data Libraries (TENDL) merges the TALYS nuclear model with data available in the JENDL \cite{nakagawa1995japanese} and ENDF \cite{chadwick2011endf} databases. In the case of $^{13}$C($\alpha, n)^{16}$O, TALYS can predict partial cross sections of different reaction channels noted as $^{13}$C($\alpha, n_{j}$)$^{16}$O, where $j$ identifies neutrons associated with different states of $^{16}$O. The TALYS-generated cross sections as a function of $\alpha$-particle energy are shown in Fig.~\ref{fig:cross_13C}. Although such a statistical model can be imprecise when predicting the detailed resonance structure as pointed out by Ref.~\cite{Febbraro2020}, its overall agreement can be used to approximate the reaction rate, allowing neutron background predictions for low-background experiments. It is reasonable to use this approach especially when precise measurements are sparse over the entire range of $\alpha$ energies relevant for ($\alpha, n$) backgrounds, which is typically broad.

In this paper, we report an analysis of several years of calibration data taken by the \mjd\, experiment, which resulted in a measurement of characteristic 6129-keV photons emitted following the $^{13}$C($\alpha,n_2)^{16}$O reactions, where the second excited state (3$^{-}$) of $^{16}$O is populated. We compare the measurement with a prediction from NeuCBOT (Neutron Calculator Based On TALYS)~\cite{westerdale2017radiogenic,ajaj2019search}. Section~\ref{sec:level2} of this paper discusses the ($\alpha,n$) reactions within the calibration sources of the \demo. Section~\ref{sec:level3} introduces the experimental techniques and analysis used to identify the 6129-keV photons. Section~\ref{sec:final_comp} describes how the TALYS-based NeuCBOT and a \textsc{Geant4}-based software for the \demo\ are used to predict the number of observable events. Section~\ref{sec:backgrounds} discusses how the same procedure can be used to estimate the background contribution to $0\nu\beta\beta$ measurements. The last section shows how similar techniques could play an essential role in future experiments with more stringent background goals.

\section{\label{sec:level2}
$^{13}$C($\alpha,n$)$^{16}$O reactions in Calibration}

\begin{figure}[htb]
 \includegraphics[width=\linewidth]{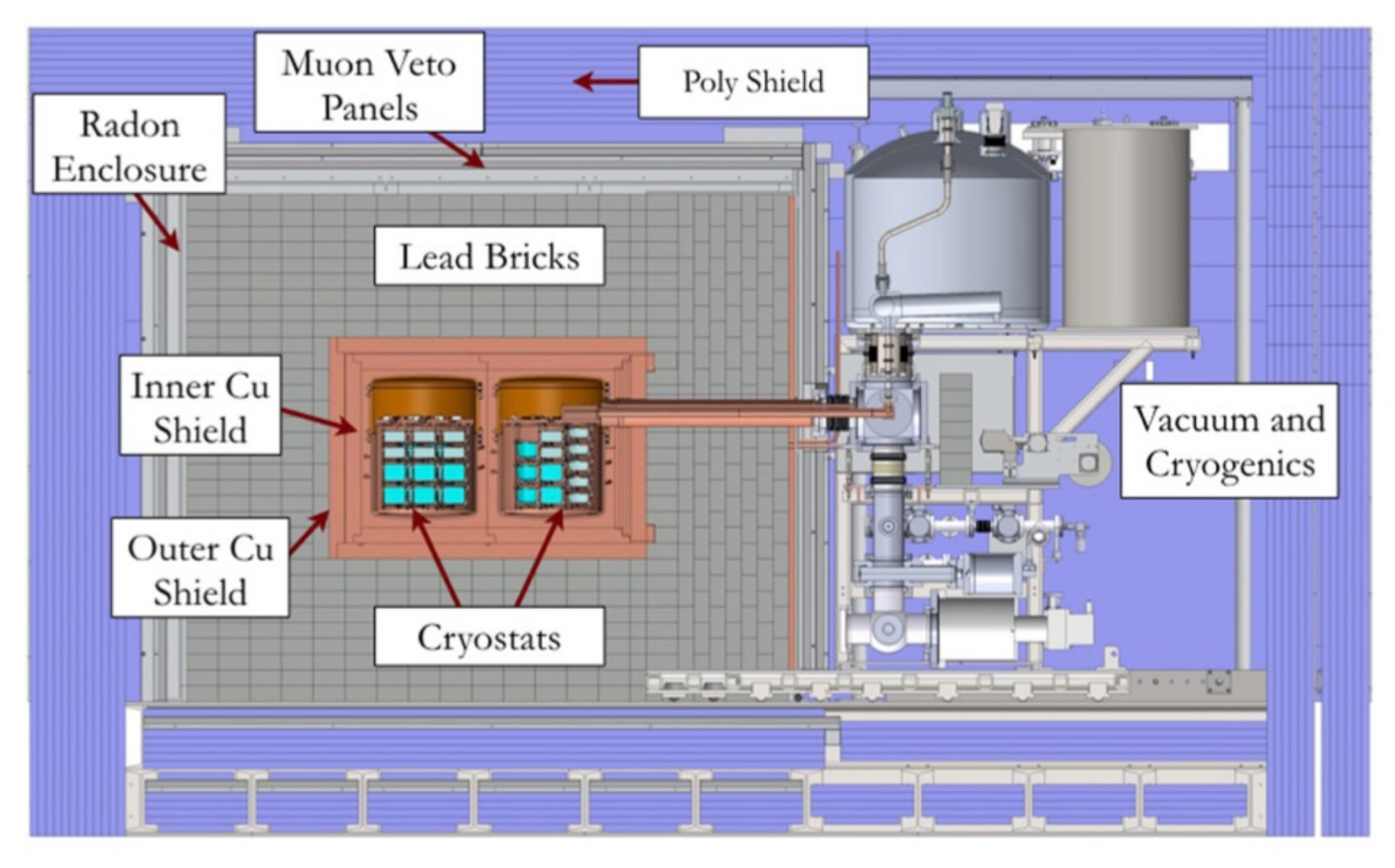}
 \caption{A schematic of the \mjd\ with two modules of HPGe detectors surrounded with layers of shielding~\cite{arnquist2021search}.}
 \label{fig:mj}
 \end{figure}

 The \mjd\ experiment searched for $0\nu\beta\beta$ in $^{76}$Ge using P-type Point Contact (PPC) High Purity Germanium (HPGe) detectors
\cite{aalseth2018search}. The \demo\ was operated at the 4850-foot level of the Sanford Underground Research Facility in Lead, South Dakota, with two modules of HPGe detectors placed in an ultra-clean and heavily shielded environment as shown in Fig~\ref{fig:mj}. The HPGe detectors had a combined total mass of 44.1 kg, of which 29.7 kg was enriched to 88\% in $^{76}$Ge with the rest being natural Ge. In March 2021, the \demo\ completed its data-taking campaign with enriched detectors and it continues taking data with natural detectors for background studies and other physics. The \textsc{Demonstrator}'s HPGe detectors in combination with low-noise electronics have achieved good linearity over a broad energy range~\cite{abgrall2020adc} and best-in-field energy resolution with a full width at half maximum (FWHM) approaching 0.1\% at the 2039~keV Q$_{\beta\beta}$ of $^{76}$Ge~\cite{alvis2019search}. This excellent energy performance coupled with the low energy threshold and low-background of the \demo\ makes it a competitive $0\nu\beta\beta$ experiment and allows for other physics beyond the Standard Model \cite{arnquist2021search,alvis2018first,abgrall2017new,abgrall2016search,alvis2019search_1}.

\begin{figure}[htbp]
 \includegraphics[width=\linewidth]{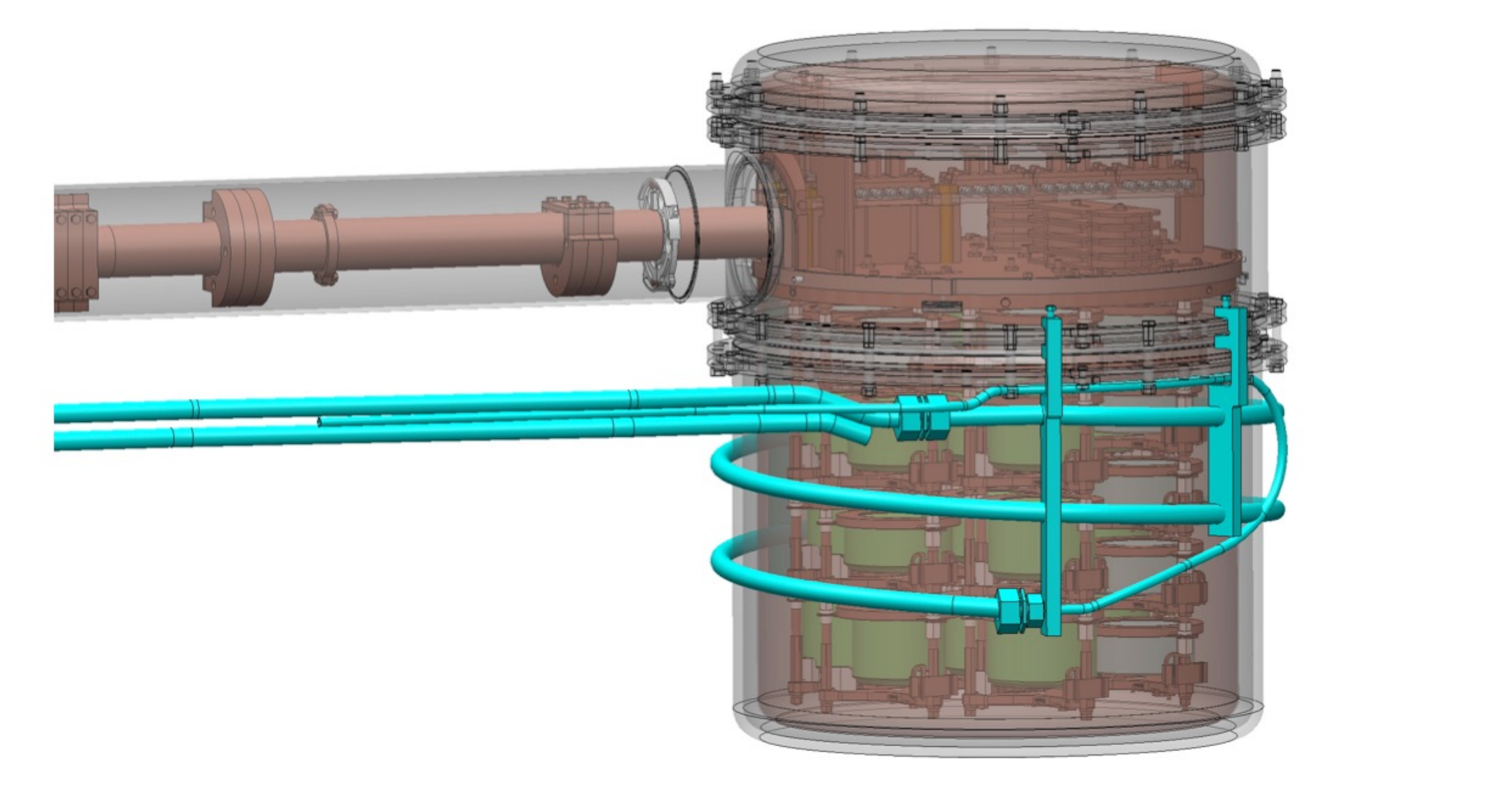}
 \caption{A diagram that shows one module, the detector strings within, and the calibration track (highlighted) through which a line source is deployed during calibrations.}
 \label{fig:track}
 \end{figure}

Ultra radiopure materials were used in the construction of the \textsc{Demonstrator}, particularly in the vicinity of germanium detectors, which are placed inside layers of compact shielding~\cite{abgrall2014majorana}. A weekly calibration is required to monitor detector stability and provide data for developing analysis cuts. The thorium isotope $^{228}$Th was selected as the calibration source because its decay chain emits several $\gamma$-rays that span from a few hundred keV up to 2615~keV, covering the Q$_{\beta\beta}$ of $^{76}$Ge and allowing for analysis over a wide energy range. The \textsc{Demonstrator}'s calibration line sources were manufactured by Eckert \& Ziegler Analytics, Inc \footnote{\url{http://www.ezag.com/home/}}. Each line source is made of thoriated epoxy encapsulated in a tube made of PTFE~\cite{abgrall2017majorana}. During calibrations, the line source was deployed into the calibration track, which surrounds the cryostat in a helical path~\cite{abgrall2017majorana}, as shown in Fig~\ref{fig:track}.  
 
The $\gamma$-rays emitted within the thorium decay chain are used for calibration and detector characterization. The decay chain ends when it reaches $^{208}$Pb producing several $\alpha$-emitters along the way. Table~\ref{tab:alpha_table} shows the energies of the main $\alpha$-particles, which lie between 5.34 MeV and 8.79 MeV. When traversing the epoxy in the calibration source, an $\alpha$-particle could initiate $(\alpha,n)$ reactions in $^{13}$C, $^{17}$O, $^{18}$O, $^{35}$Cl, and $^{37}$Cl, of which reactions with $^{13}$C dominate. $^{13}$C($\alpha,n_{2})^{16}$O reactions are possible with $\alpha$-particles above about 5~MeV, resulting in 6129-keV photons. The \textsc{Demonstrator}'s excellent energy performance allowed a clear observation of this 6129-keV signature on top of the thorium photon energy spectrum during calibrations. 

\begin{figure}[htb]
 \includegraphics[width= \linewidth]{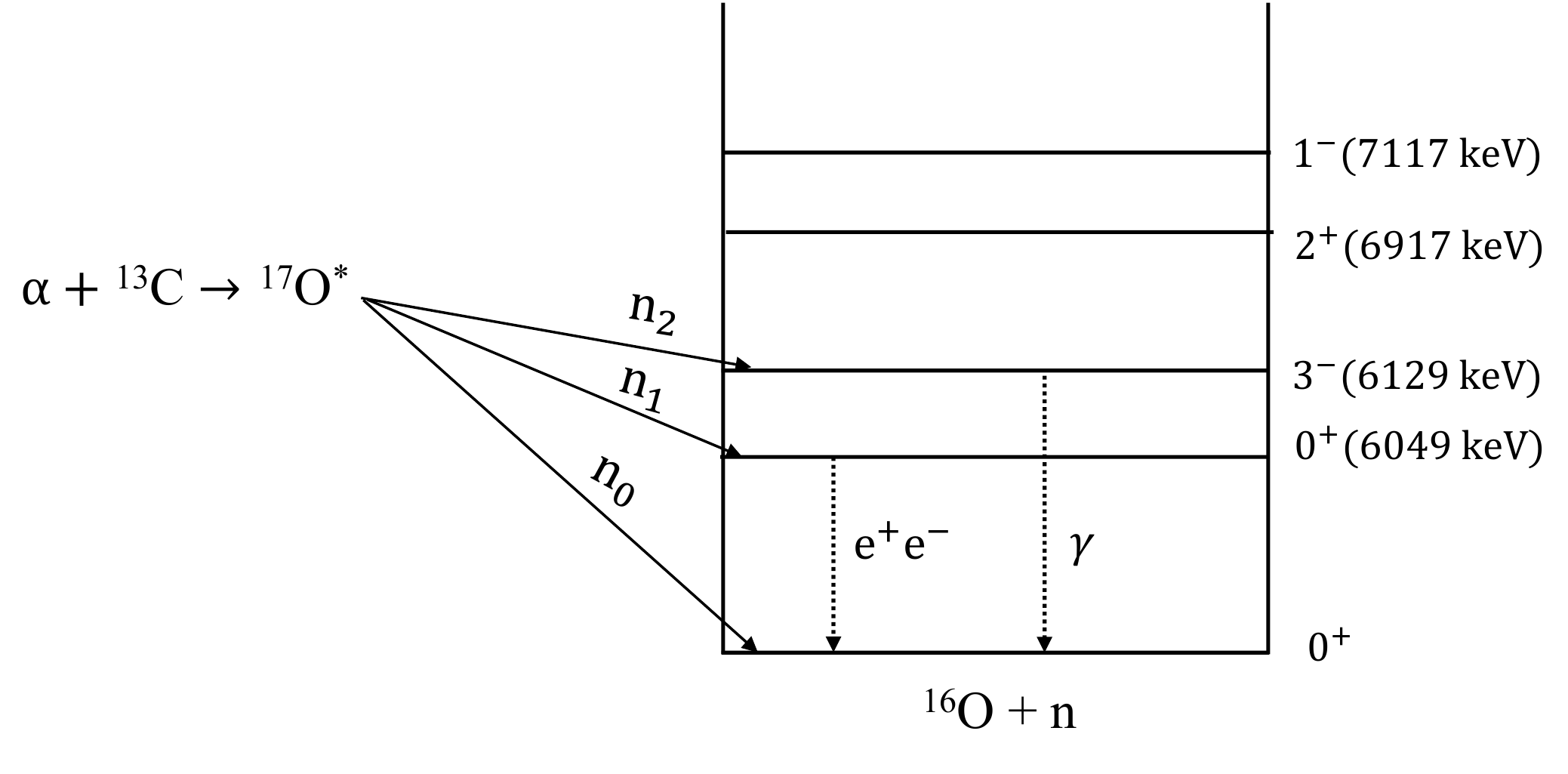}
 \caption{The level scheme of $^{16}$O as populated in the $^{13}$C$(\alpha,n)^{16}$O reaction (energy not to scale)  simplified from Figure~1 of Ref.~\cite{Febbraro2020}. The numerical index of the emitted neutrons n$_{0}$, n$_{1}$, n$_{2}$ represents which final state in $^{16}$O is populated. Due to selection rules, the 0$^{+}$ (6049~keV) state deexcites via the emission of an e$^{+}$e$^{-}$ pair, while the 3$^{-}$ (6129~keV) state deexcites through $\gamma$-ray emission. Data from \cite{mohr2018revised,Febbraro2020}.}.
 \label{fig:rxn}
 \end{figure}

 In $^{13}$C($\alpha,n$)$^{16}$O reactions, an $\alpha$-particle is captured in $^{13}$C to form the compound nucleus $^{17}$O$^*$, which decays to the ground state or excited states of $^{16}$O by emitting a neutron. Figure~\ref{fig:rxn} shows the simplified level scheme of $^{16}$O that can be populated from the decay of $^{17}$O$^{*}$. Since the $\alpha$-particles in the thorium chain have energy up to 8.79 MeV as listed in Table~\ref{tab:alpha_table}, they can potentially open up the reactions channels of ($\alpha,n_{1}$), ($\alpha,n_{2}$), ($\alpha,n_{3}$), and ($\alpha,n_{4}$). The population of the second excited state (3$^{-}$) of $^{16}$O at 6129 keV is favored for $\alpha$-particles with energy above 6~MeV, as shown by the turquoise line in Fig.~\ref{fig:cross_13C}. The isomeric transition of the (3$^{-}$) state to the ground state of $^{16}$O emits a characteristic 6129-keV photon, presenting a unique signature to look for in calibration data. The $^{13}$C($\alpha,n_2$)$^{16}$O reaction is described in Eq.~\ref{eq:13c}.
  
\begin{equation}
\begin{split}
       ^{13}\rm{C} + \alpha & \rightarrow ^{17}\rm{O}^{*} \\
       & \rightarrow ^{16}\rm{O}^{*} (3^{-}) + n \\
       & \rightarrow ^{16}\rm{O}(g.s.) + \gamma~(6129~\rm{keV}) + n
       \label{eq:13c}
    \end{split}
   \end{equation}

\begin{table}[htpb]

\caption{Energies of primary $\alpha$-particles from the decay chain of $^{228}$Th. Energy data are taken from Nuclear structure \& decay Data (NuDat 3.0)\footnote{\url{https://www.nndc.bnl.gov/nudat3/}}.}
    
\label{tab:alpha_table}
    
\centering
    
\begin{tabular}{|c | c| c|}
        \hline
\thead{$\alpha$-particle energy\\ (MeV)}& Parent isotope & \thead{Intensity \\ (per $^{228}$Th decay)} \\ [0.5ex] 
\hline 
 5.423 &$^{228}$Th & 0.734    \\
 \hline
 5.340 &$^{228}$Th & 0.260    \\
\hline
 5.685 &$^{224}$Ra  &0.949  \\
\hline
5.449 &$^{224}$Ra  &0.051  \\
\hline
 6.288 &$^{220}$Rn  & 0.999   \\
\hline
 6.778 &$^{216}$Po  &0.999  \\
\hline
 6.050 &$^{212}$Bi  &0.090 \\
 \hline
 6.089 &$^{212}$Bi  &0.035 \\
\hline
 8.785 &$^{212}$Po  &0.641   \\
\hline
\end{tabular}

\end{table}

\section{\label{sec:level3}Analysis of data}
For this work, data from weekly calibrations was analyzed in several steps, including data selection and data quality checks, validation of the \textsc{Geant4} simulation, and the signature search at higher energies. The \mjd\ Data Acquisition (DAQ) system records waveforms from each HPGe detector using two digitization channels with different amplifications, called the low-gain and high-gain channels. The high-gain channels have been extensively used for double-beta decay searches \cite{alvis2019search}, but they saturate around 3~MeV due to the digitization range. 
The low-gain channels have a wider dynamic range up to 10~MeV and allow a study of signatures with higher energy depositions, \textit{e.g.} by cosmic ray reactions or neutrons. The low-gain channels are used here to search for the 6129-keV photons.


\subsection{\label{sec:bench} Data quality and simulation benchmarking}
The modular approach of the \demo\ enabled a flexible construction as well as early data-taking once the first module was constructed. Each calibration source was deployed separately for most of the \textsc{Demonstrator}'s calibration data, except for a period after the installation of the second module when two sources were deployed simultaneously to calibrate both modules. For these calibrations, the DAQ throughput was potentially saturated. Thus this analysis only uses data collected when one calibration source was deployed at a time. 
Due to evolving calibration procedures, early commissioning data are not used. For example, during commissioning, transition runs during which the source was in motion were not flagged, which created larger uncertainties in analysis time boundaries. The data analyzed here include calibration data sets from the years 2016-2019, which were also used in the analysis of the recent double-beta decay results~\cite{alvis2019search}. The data quality checks and channel selection used in the double-beta decay analysis~\cite{alvis2019search} were also applied here. Additional data quality checks based on the prominent 2615-keV $\gamma$-peak following the $\beta^{-}$ decay of $^{208}$Tl are applied to the calibration data used in this analysis. 
If the 2615-keV, full energy event rate in a run is found to deviate more than 3.5 $\sigma$ from the mean rate in the same data set, the run is excluded from this analysis. Such deviations can occur when, for example, the nitrogen dewars were filled, since the flow of liquid nitrogen induced noise.

After the data quality checks, we compared the observed source activity ($A_{observed}$) with the expected activity ($A_{expected}$), defined as:

\begin{equation}
    A_{observed} = \frac{R}{\epsilon \times b}
    \label{eq:obs}
\end{equation}

\begin{equation}
    A_{expected} = A_{0} e^{[-\lambda (t-t_{0})]}
    \label{eq:exp}
\end{equation}
In Eq.~\ref{eq:obs}, the observed activity of a calibration source during each weekly calibration was estimated based on the rate, $R$, of the full energy 2615~keV peak, the corresponding efficiency, $\epsilon$, of detecting the full energy 2615-keV photons, and the branching ratio, $b$, for the $^{212}$Bi$\rightarrow^{208}$Tl transition in the $^{228}$Th decay chain. 
The \textsc{Geant4}-based \cite{agostinelli2003Geant4} simulation package, \textsc{MaGe}~\cite{boswell2011mage}, was used to estimate the detection efficiency ($\epsilon$) of the 2615-keV photons originating from the calibration sources in their deployed positions. In Eq.~\ref{eq:exp}, the expected activity of each calibration source is projected for every weekly calibration based on the initial activity, $A_{0}$, reported by the vendor at a given time $t_0$, the decay constant, $\lambda$, and the time of each calibration, $t$. The decay chain is in equilibrium, so the decay constant is based on the 1.9-year half-life of $^{228}$Th. Uncertainties in the branching ratio, decay constant, and calibration time are negligible, so the uncertainty in the expected activity is dominated by the uncertainty in $A_0$, which was $10.36 \pm 0.60$ kBq on May 1, 2013.

\begin{figure}[htpb]
  \includegraphics[width=\linewidth]{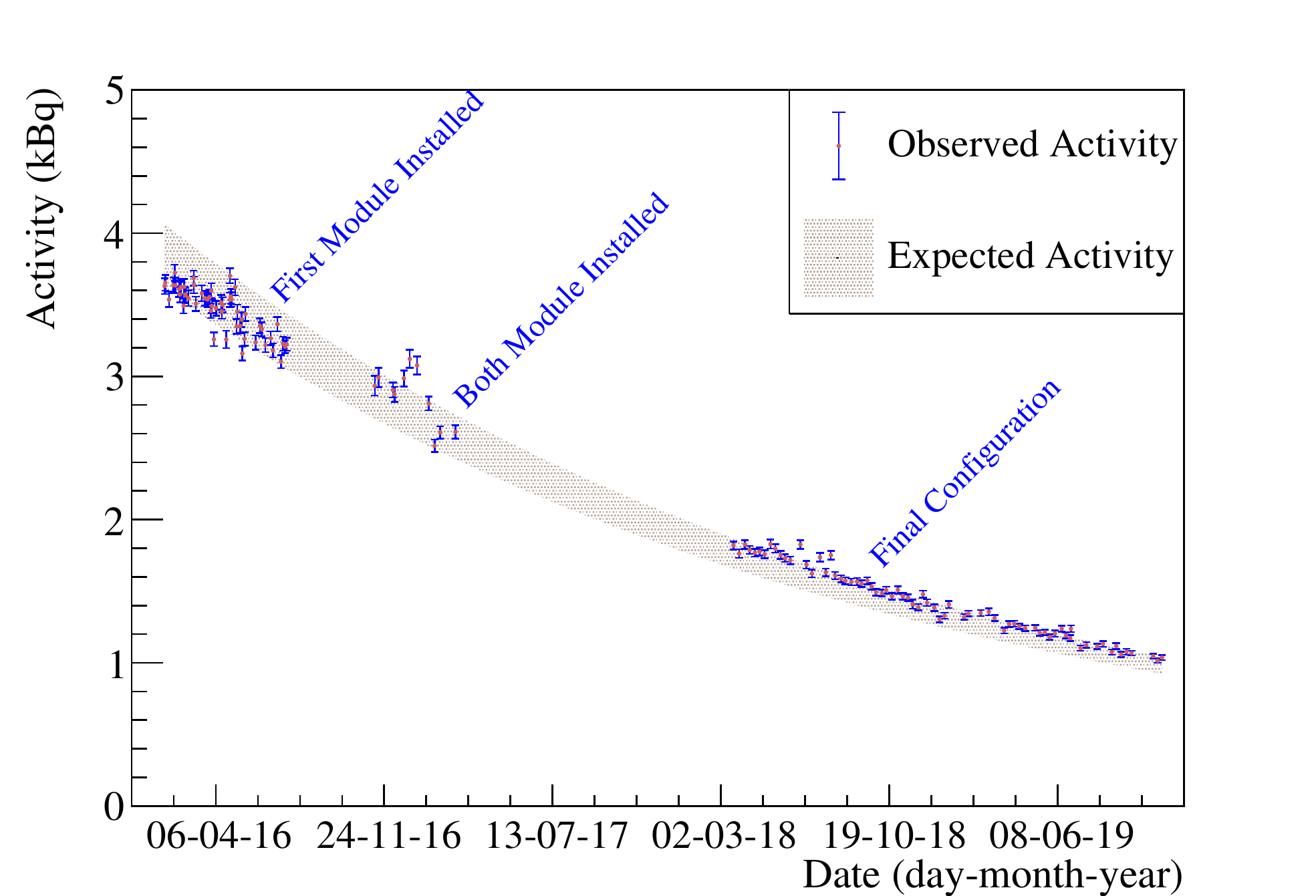}
  \includegraphics[width=\linewidth]{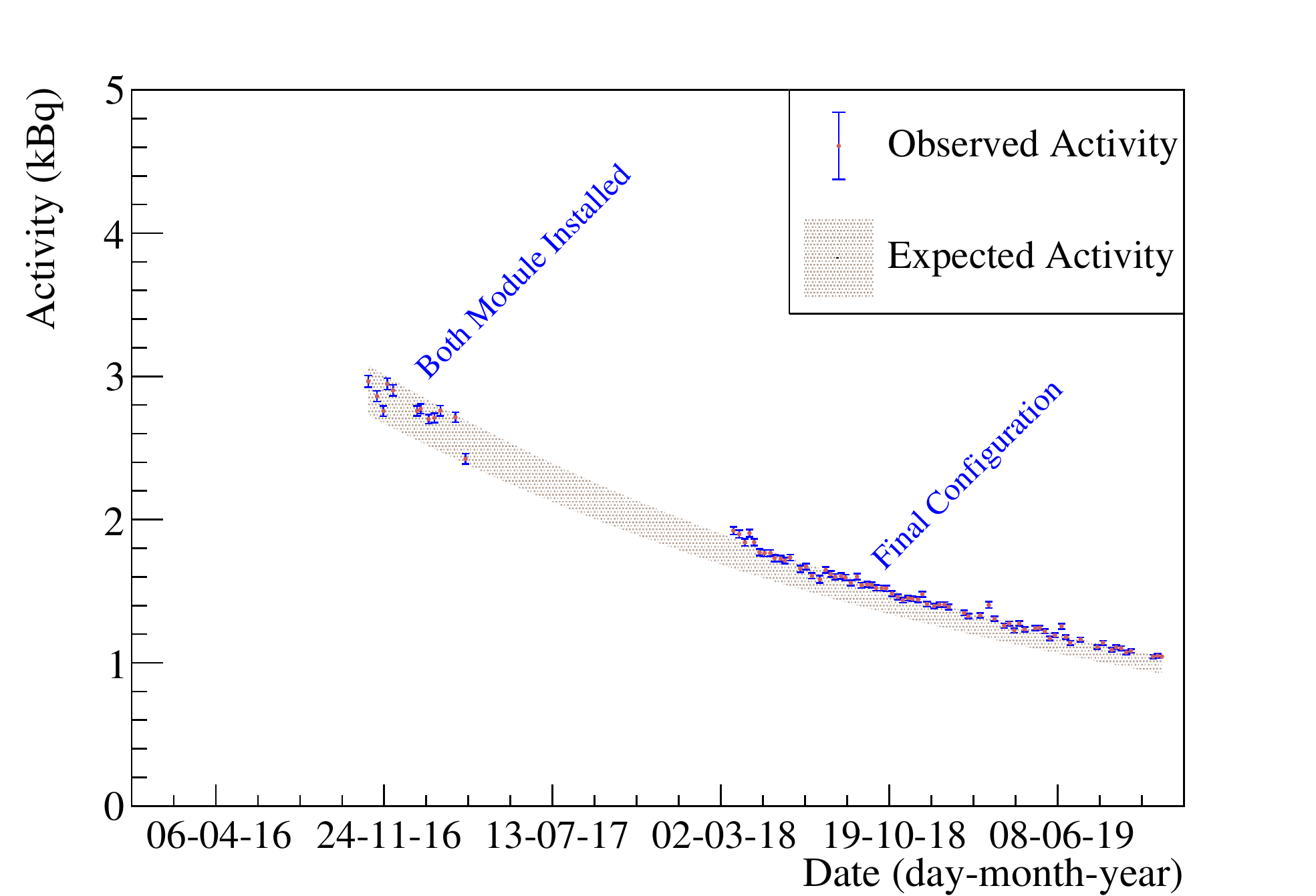}
  
 \caption{Observed and expected activities for the two source assemblies used in \mjd. The data points indicate the observed activity of each source assembly for each weekly calibration, while the band represents the expected activity which includes the vendor reported uncertainty. The uncertainties in the observed activity are statistical only.}
 \label{fig:act}
 \end{figure}

As shown in Fig.~\ref{fig:act}, a good agreement was found between the expected and the observed activity over multiple years of calibration data for both source assemblies. This implies good accuracy for the simulations performed by \textsc{MaGe} and gives confidence that \textsc{MaGe} can make correct efficiency predictions for the analysis of the 6129~keV $\gamma$-rays.

\subsection{Signature search}
\begin{figure}[htpb]
 \includegraphics[width=\columnwidth]{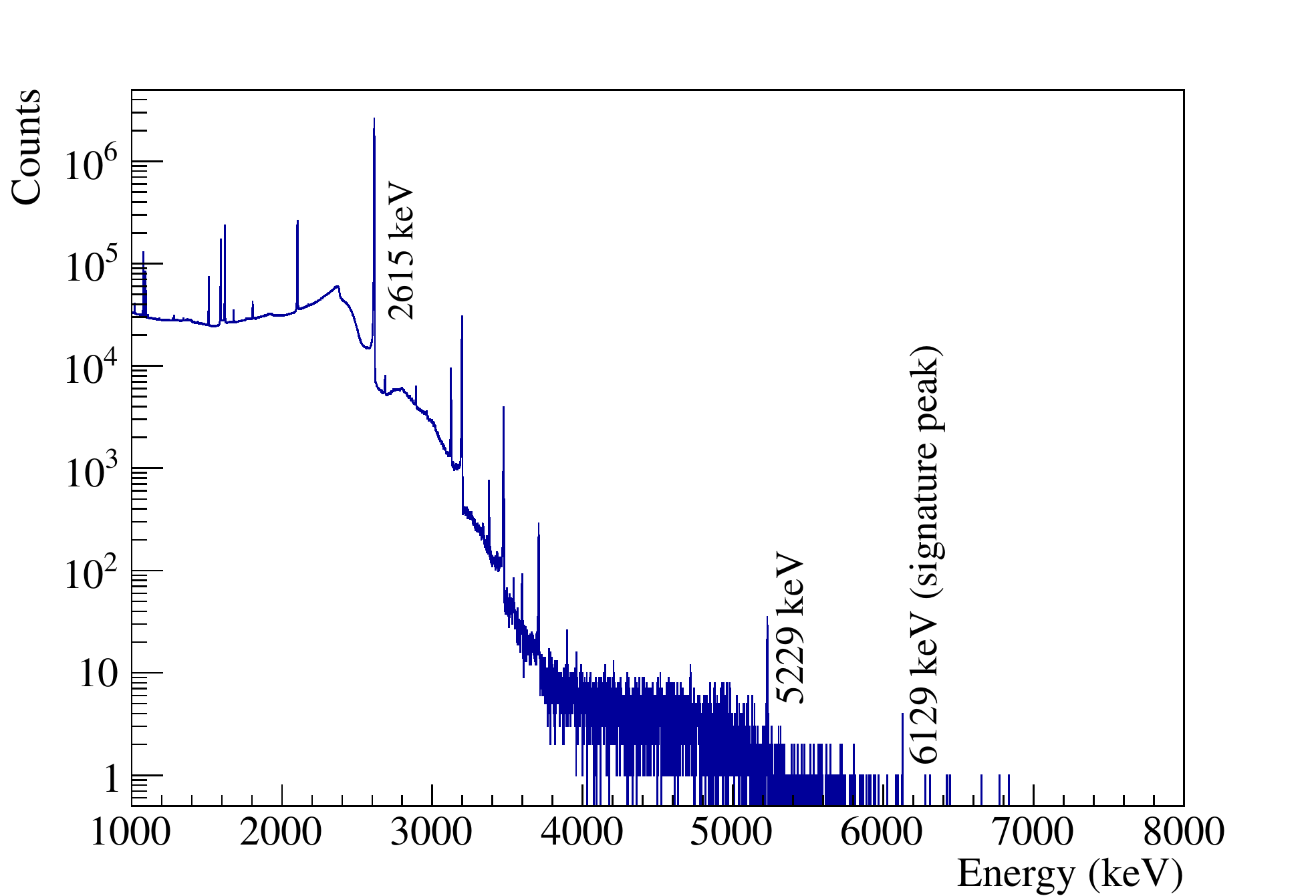}
 \caption{The sum energy spectrum of calibration data selected for the analysis. It shows various $\gamma$-ray peaks, including 2615~keV, the signature peak of 6129~keV, and other peaks from the calibration source and peaks due to random coincidence events, and summing. }
 \label{fig:spec}
 \end{figure}

\begin{figure}[htb]

 \includegraphics[width=\linewidth]{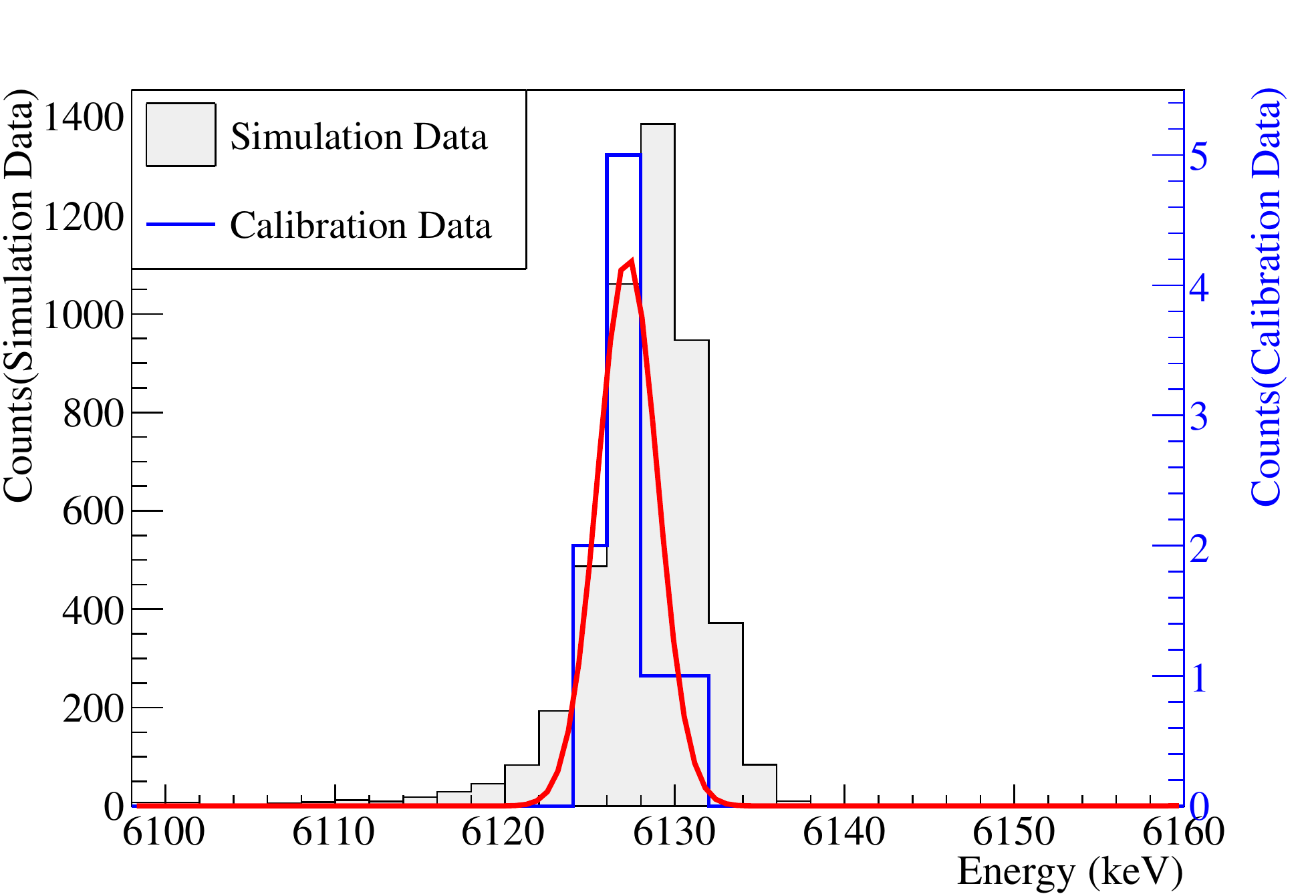}
 \caption{The signature peak at 6129~keV from $^{13}$C($\alpha,n_{2}$)$^{16}$O reactions in the \mjd\ calibration sources, shown in blue color and fitted with Gaussian in red. The gray-filled histogram is the peak shape from the simulation of 1 million 6129-keV photons from the calibration tracks.}
 \label{fig:sig}
 \end{figure}

\begin{figure}[htbp]
 \includegraphics[width=\linewidth]{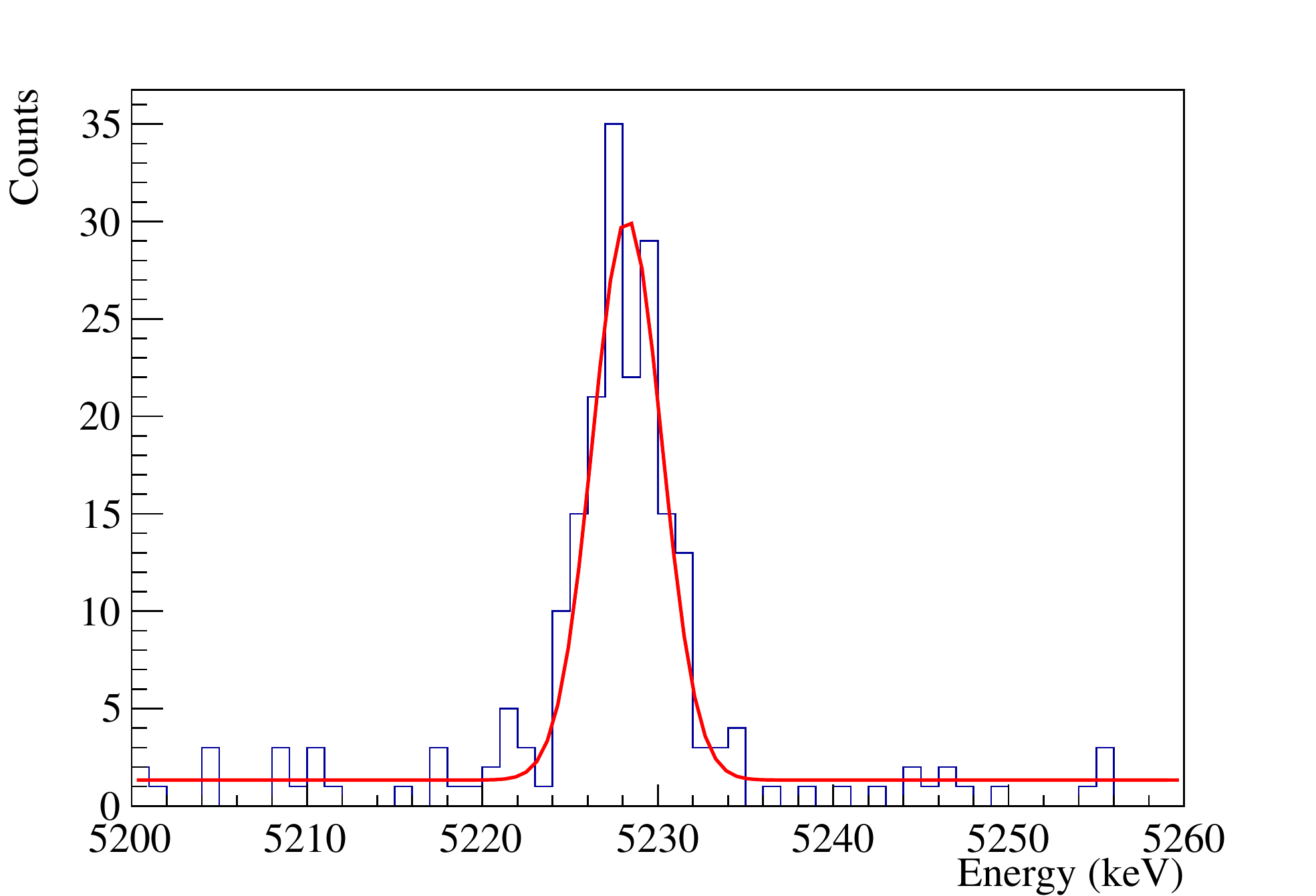}
 \caption{The double coincidence peak at 5229 keV ($2615.5$~keV$\times2$) in the sum energy spectrum of the calibration data is fitted with with a Gaussian plus a flat background.}
 \label{fig:double}
 \end{figure}

The search for the 6129-keV photons from the $^{13}$C($\alpha,n_{2}$)$^{16}$O reactions was performed using the sum energy of events, which is obtained by summing all coincident energy depositions over all active HPGe detectors within a 4~$\mu$s window~\cite{arnquist2021search}. This sum energy is used because of the high probability for several-MeV photons to distribute their full energy in multiple detectors. Fig.~\ref{fig:spec} shows the sum energy spectrum above 1~MeV in calibration data. 
The signature at 6129~keV following the $^{13}$C($\alpha,n_{2}$)$^{16}$O reaction is clearly visible. Fig.~\ref{fig:sig} provides a spectrum in a smaller energy band around the 6129~keV region.
Events above 2615~keV are mostly due to summing, or random coincidences of two unrelated decays in the calibration source. For the latter one, the most prominent feature is the 5229-keV peak. When two 2615-keV photons, the energy of which is 2614.511~keV, are in coincidence, the sum energy appears to be twice of a photon energy. The zoomed-in plot of this peak is shown in Fig.~\ref{fig:double}.

We defined the region of interest (ROI) for the 6129-keV peak search as (6129 $\pm$ 10)~keV based on the expected resolution in that energy region: about 2~keV ($1\sigma$) at 6~MeV, so the chosen window covers about 5$\sigma$ on each side of the peak. A simple Gaussian fit to the signal peak found the mean to be 6127$\pm$0.6 keV and the standard deviation to be 1.8$\pm$0.4 keV, as shown in Fig.~\ref{fig:sig}. A total of 9 events were found in the ROI in all data combined. Given the low statistics, the uncertainties from the fit are relatively large and less robust. 
As a cross-check, a simple Gaussian plus a flat background was fit to the much stronger double coincidence 5229-keV peak in Fig.~\ref{fig:double}, where the mean was found to be 5228$\pm$0.2 keV with a standard deviation of  2.0$\pm$0.1 keV. These full energy peaks are seen at their expected locations and with their expected widths in the sum energy spectrum from the low-gain channels, indicating a great energy performance extended to the energy range of multiple-MeV. 

As seen in Fig.~\ref{fig:sig}, the signature peak at 6129~keV stands out clearly, so all of the 9 observed events in the peak are considered to be the signal, \textit{i.e.} 6129-keV photons following the $^{13}$C($\alpha,n_2$)$^{16}$O reactions. Given that no background events were found for at least 20 keV on both sides of the peak outside the ROI, the potential background in the 40-keV region from 6099 keV to 6159 keV excluding the 20-keV ROI can be determined as at most 1.29 counts at a $1\sigma$ level, which translates to a $1\sigma$ upper limit of 0.64 counts of background in the ROI. To better determine the potential background contribution, we also counted events in a much broader background region from 6 to 6.5~MeV, excluding the ROI around the 6129-keV peak. Based on 8 events in this 480-keV background region, we projected the potential background to be 0.33 counts in the ROI. Incidentally, this projects 0.67 counts of background from 6099 keV to 6159 keV excluding the 20-keV ROI, statistically consistent with observing none, which would happen with a 50\% probability. In short, the observed number of signal events in the ROI in the combined data sets is 9, while 0.33 counts is the estimated background contribution to the expected number of events. The difference between 0.33 counts and 0.64 counts is treated as a systematic uncertainty on the background contribution to the ROI.

\section{\label{sec:final_comp}
Comparisons with predictions}
\subsection{Prediction calculation}
NeuCBOT~\cite{westerdale2017radiogenic} is a software tool based on TALYS to calculate the neutron yield and neutron energy spectra for ($\alpha,n$) reactions in materials. It models the entire trajectory of $\alpha$-particles: initializing $\alpha$-particles according to ENSDF evaluated nuclear decay data \cite{tuli1996evaluated}, tracking their energy loss and range according to SRIM \cite{ziegler1985stopping}, and ultimately predicting the ($\alpha,n$) rate based on cross sections in TALYS-based TENDL. In this work, the 6129-keV photon production rate was estimated by NeuCBOT using the partial $^{13}$C($\alpha,n_{2}$)$^{16}$O cross sections from TALYS-1.95, and it was found to be 2.98 $\times 10^{-7}$ $\gamma$/Th-decay. 

The detector configuration, such as the list of active detectors, can vary over time. The source activity also reduces as thorium decays away. Therefore, the number of predicted events was calculated for each weekly calibration and summed together using:
\begin{equation}
 N = Y \times \sum_{i} A_{i} \times \epsilon_{i} \times T_{i}
 \label{eq:final_expec}
\end{equation}
 Here, $Y$ is the $\gamma$-ray production yield per decay of thorium, which is constant for all data sets since the source assembly does not change. For each weekly calibration $i$, the factors A$_{i}$, $\epsilon_{i}$, and T$_{i}$ are the source activity, detection efficiency for the 6129-keV photons, and live time, respectively. The efficiency, $\epsilon_{i}$, was calculated with \textsc{MaGe} for the 6129-keV photon using the same geometry as for the 2615-keV analysis, but the sum energy was used instead of individual detector energy for consistency. The simulated 6129-keV peak shape in the sum energy spectrum for 6129-keV photons uniformly seeded inside the calibration source is shown in Fig.~\ref{fig:sig}. The same ROI as in the data analysis was used to calculate the efficiency. Realistic energy responses, including dead layer models of each detector~\cite{alvis2019search}, are folded into the simulation, so the peak has slight deviations from Gaussian, notably a low energy tail. 
 
 \subsection{Comparison of observed and expected events}
Table~\ref{tab:final_table} compares the expected number of 6129-keV $\gamma$-rays with the number observed; the latter can be modelled by Poisson statistics with an unknown true mean. Based on the observed signal counts, the confidence interval on the mean of Poisson signals is calculated at a 90\% confidence level (C.L.) using the Feldman-Cousins statistical approach for small signals \cite{feldman1998unified}. 

Sources of uncertainty in the expected counts are summarized in Table~\ref{tab:unc}. Uncertainties in the SRIM database are reported in Ref.~\cite{heaton1989neutron}. Uncertainties due to the chemical composition of the epoxy material and in the source activities were both based on the specifications provided by the vendor. 
As discussed before, the projected background contribution in the ROI depends on the choice of background regions and the difference between the narrow 40-keV and the wide 480-keV background regions is taken as the uncertainty. To assess uncertainty associated with the calculation of the high energy photon detection efficiency using the sum energy, we repeated the calibration source activity analysis in Section \ref{sec:level3} using the sum energy. On average, a 11.9\% difference in the source activity is observed between calculations based on single detector energy and the sum energy at 2615-keV. The difference between the vendor specification and the source activity based on the sum energy was found to be smaller, so the 11.9\% is an overestimation of the systematic uncertainty in simulation.


\begin{table*}[htpb]
\caption{Expected and observed counts of 6129-keV photons. Expected counts are estimated based on Eq.~\ref{eq:final_expec}, and the corresponding uncertainties are the combination of various uncertainties shown in Table~\ref{tab:unc}. The range of signal mean is the 90\% C.L. interval of Poisson signal mean based on observed signal counts in each data~\cite{feldman1998unified}. The individual data set labeling follows \demo\ configuration changes as explained in Ref.~\cite{alvis2019search}.}
 
\label{tab:final_table}
 
\centering
\begin{tabular}{|c |c |c | c| c|}
  \hline
 Calibration Data Set & \thead{Integrated Exposure Time \\(hour)} &Expected Counts & Observed Counts & \thead{90\% Interval of Signal Mean \\ given Observation}\\ [0.5ex] 
\hline 
 DS1 &40.2 &0.42$\pm$0.07 &0 &[0.00,~2.44] \\

\hline
 DS2 &13.4 &0.13$\pm$0.02 &1 &[0.11,~4.36] \\

\hline
 DS5 &41.8 &0.41$\pm$0.07 &1 &[0.11,~4.36] \\

\hline
 DS6a &43.9 &0.32$\pm$0.05 &1 &[0.11,~4.36] \\

\hline
 DS6b &178.3 &1.19$\pm$0.20 &4 &[1.47,~8.60] \\

\hline
 DS6c &245.0 &1.27$\pm$0.21 &2 &[0.53,~5.91] \\

\hline
 Total &562.6 & 3.74$\pm$0.63 &9 &[4.36,~15.30]\\

\hline
\end{tabular}

\end{table*}


\begin{table*}[htpb]
\caption{Relative uncertainties for the expected number of counts. The total systematic uncertainty is the sum in quadrature of individual systematic contributions.}
 \centering
 \begin{tabular}{c c}
 
  \hline
   $\gamma$ yield value due to uncertainties in the SRIM reported in ~\cite{heaton1989neutron} & 5.0\%\\
   Chemical composition in epoxy & 4.0\% \\
   Activity of the source as reported by Eckert \& Ziegler & 5.8\% \\
    Systematic uncertainty in simulation& 11.9\%
  \\ 
   Statistical uncertainty in simulation & 1-2 \% (neglected)\\
   Systematic uncertainty in background contribution & 8.3\% \\   
    Total systematic uncertainty & 16.9\% \\
    \hline
 \end{tabular}
 \label{tab:unc}
\end{table*}

\begin{figure}[htp]


\includegraphics[width=\linewidth]{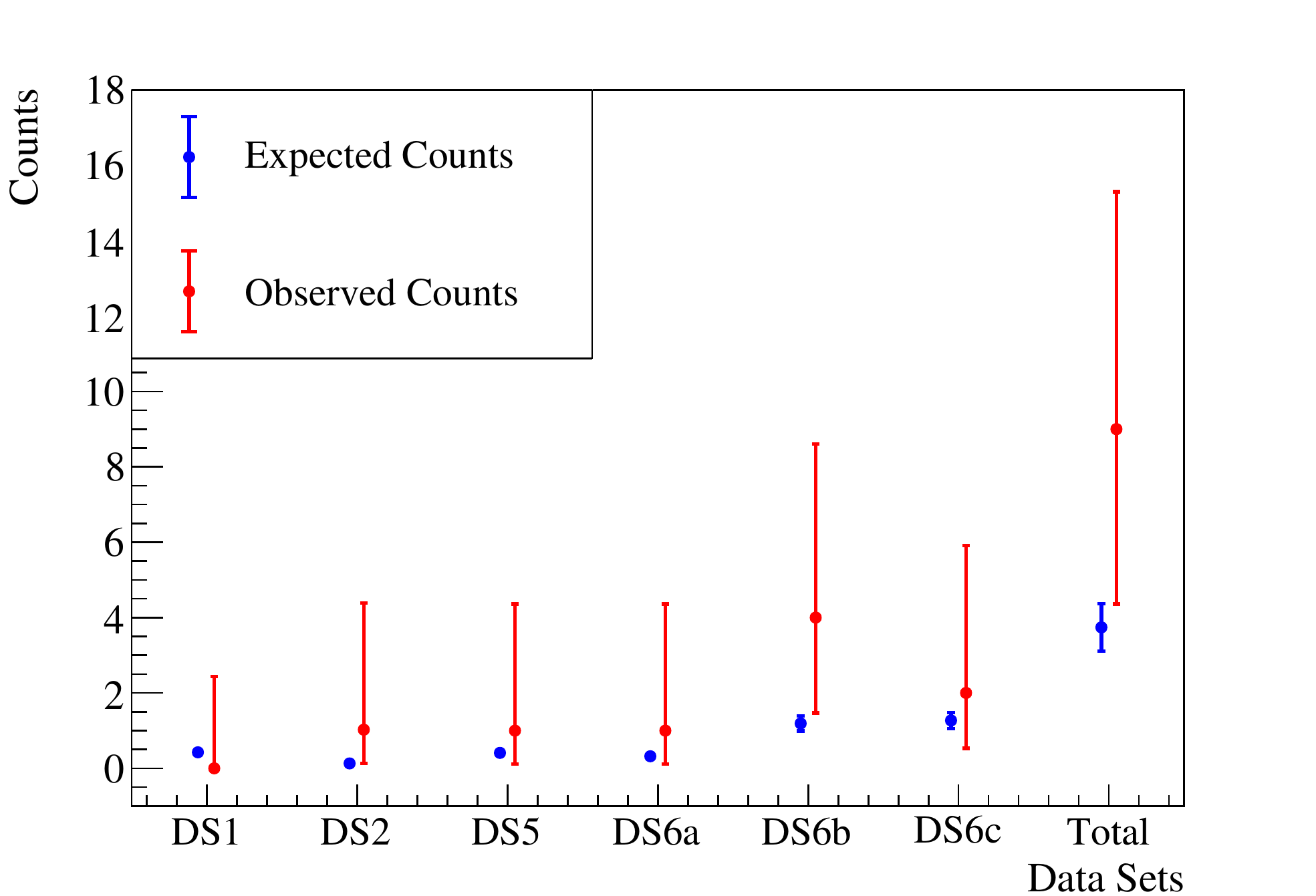}
 
 \caption{Expected and observed number of 6129-keV photons with corresponding uncertainties in each data set and in the combined data set. The error bars in the observed counts indicate the 90\% C.L. intervals on the mean of Poisson signals as listed in Table~\ref{tab:final_table}.}
 \label{fig:final}
 \end{figure}

Figure~\ref{fig:final} visualizes the comparison between the expected and the observed number of 6129-keV photons. 
The observed number of events tends to be higher than the expected number, however the statistical uncertainty in the experimental data is large. The range of expected counts is overall consistent with the 90\% confidence interval on the observed signal strength. This comparison suggests that TALYS cross sections combined with SRIM enables reasonable estimations of ($\alpha,n$) rates. This consistency at a 90\% confidence level lends support to the approach of predicting neutron production from $\alpha$-induced reactions in low-background experiments using the presented tools.


\section{\label{sec:backgrounds}Background estimation for $0\nu \beta\beta$ search}
Neutrons produced by ($\alpha,n$) reactions during the \textsc{Demonstrator}'s calibration runs can enter the germanium crystals and get captured. After each calibration, these sources were retracted to parked locations entirely outside the shield. Therefore, only the neutrons produced during the calibration are of concern. When $^{76}$Ge undergoes neutron capture, the ground state of $^{77}$Ge or the metastable state, $^{77m}$Ge, can be produced, both of which could $\beta$ decay with energy releases larger than the 2039~keV Q$_{\beta\beta}$ of $0\nu\beta\beta$ in germanium~\cite{arnquist2022signatures}. 
The main background contributor here is the long-lived isotope $^{77}$Ge with a half-life of 11.2 hr, which can decay during the $0\nu\beta\beta$ decay data-taking periods following the hours-long calibration periods. The metastable state $^{77m}$Ge with a 54-second half-life is less of a concern.

 \begin{figure}[htp]
 \includegraphics[width=\linewidth]{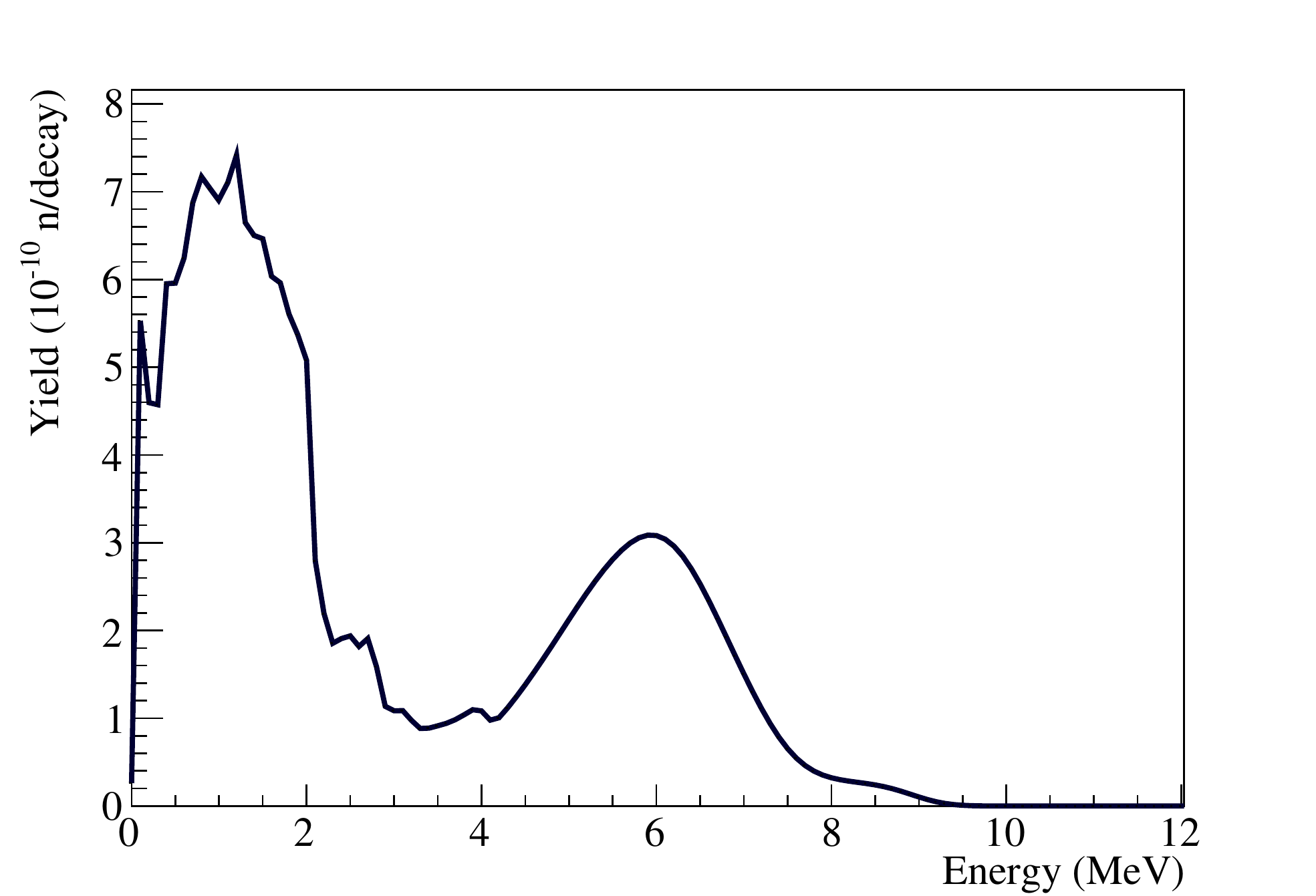}
 \caption{Neutron energy spectrum from the ($\alpha,n$) reactions in the epoxy. The spectrum is obtained by using NeuCBOT based on TALYS-1.95 generated cross sections data.}
 \label{fig:n_spec}
 \end{figure}

Fig.~\ref{fig:n_spec} shows the NeuCBOT calculation of energies and yields of neutrons generated from all types of ($\alpha$, n) reactions within the calibration sources.
\textsc{MaGe} was used to estimate the production and decay of $^{77}$Ge inside the germanium crystals given this neutron flux. This background contribution was estimated to be on the order of $10^{-5}$ cts/(keV-kg-year) before any analysis cuts. This shows that calibration neutrons are a negligible contribution compared with the total background measured in the \demo\ \cite{alvis2019search}.

The GERDA experiment investigated a similar background source in their Phase I data taking~\cite{baudis2015production}. They estimated a background contribution of $10^{-4}$ cts/(keV-kg-year) for $0\nu\beta\beta$ by neutrons from calibration sources. The higher background index can be explained by the stronger activity and slightly different geometry used in GERDA. For GERDA Phase II data taking, this background was minimized by deploying a new design of gold-encapsulated thorium calibration source~\cite{baudis2015production}. This design reduces the possible interaction of $\alpha$-particles and it is adapted by the LEGEND calibration system~\cite{abgrall2021legend}. 

Next-generation experiments searching for $0\nu\beta\beta$ have much more stringent background requirements. Hence, potential background sources of radiogenic ($\alpha,n$) neutrons from detector construction materials should be examined carefully. 
While extensive efforts are in place to shield room and cosmogenic neutrons, some neutron sources could be inside the water shielding or are even introduced by shielding materials~\cite{arnquist2022signatures}. One example is the large steel cryostat which houses the LEGEND main argon volume. The combination of TALYS-based software can be valuable to provide rough estimations in these cases, as investigated in Refs.~\cite{abgrall2021legend,Barton2021}, using NeuCBOT in combination with \textsc{Geant4}.
\section{\label{sec:level4} Discussion and Summary}
The work presented above combines the achievements of the \mjd\ experiment in terms of excellent energy performance and robust as-built simulations. The search for signatures in a wide energy range, well beyond the Q$_{\beta\beta}$ of $^{76}$Ge, is possible due to excellent energy linearity and resolution of the \mjd. These achievements result from the intrinsic advantages of HPGe detectors in combination with low-noise electronics and dedicated efforts on energy estimation corrections and calibrations. The analysis presented here found a good agreement between detected and expected energy for the signature at 6129~keV and verified the algorithms at the sum peak of two 2615-keV $\gamma$-rays. 
We have shown that the measured rate is consistent with simulations over various detector configurations in multiple years of calibrations. The agreement between the expected decay activity and the observed activity of the calibration sources in the \mjd\ is reported for the first time, demonstrating the excellent performance of the \textsc{MaGe} simulation software, which is also used by GERDA and LEGEND.

Our work shows how signatures of ($\alpha,n$) reactions can be detected in low-background experiments and how simulations are crucial in understanding this radiogenic neutron background. The agreement between simulated and measured rate is valuable feedback since ($\alpha,n$) data can be sparse, and 
can have significant discrepancies, as pointed out by Ref.~\cite{Febbraro2020}. At 90\% C.L., our measurement of the 6129-keV photons from the second excited state in $^{16}$O is consistent with the predictions generated by the TALYS-based NeuCBOT program, although the statistical uncertainty is large. This suggests that the TALYS-based NeuCBOT provides a reasonable estimation of neutrons from thorium impurities in carbon-rich organic materials. Our findings are widely applicable, as thorium is one of the most common impurities, and carbon-rich organic materials such as various plastics and epoxies are often used in experiments in abundance. It is reasonable to expect that ($\alpha,n$) reactions induced by alpha particles from thorium impurities in a range of carbon-rich organic materials share similar profiles. While future experiments may utilize materials with higher radiopurity than the current experiments, the size and length of future experiments can result in a similar ($\alpha,n$) background contribution for these rare event searches.

\section{\label{sec:ack} Acknowledgements}
This material is based upon work supported by the U.S.~Department of Energy, Office of Science, Office of Nuclear Physics under contract / award numbers DE-AC02-05CH11231, DE-AC05-00OR22725, DE-AC05-76RL0130, DE-FG02-97ER41020, DE-FG02-97ER41033, DE-FG02-97ER41041, DE-SC0012612, DE-SC0014445, DE-SC0018060, and LANLEM77/LANLEM78. We acknowledge support from the Particle Astrophysics Program and Nuclear Physics Program of the National Science Foundation through grant numbers MRI-0923142, PHY-1003399, PHY-1102292, PHY-1206314, PHY-1614611, PHY-1812409, PHY-1812356, and PHY-2111140. We gratefully acknowledge the support of the Laboratory Directed Research \& Development (LDRD) program at Lawrence Berkeley National Laboratory for this work. We gratefully acknowledge the support of the U.S.~Department of Energy through the Los Alamos National Laboratory LDRD Program and through the Pacific Northwest National Laboratory LDRD Program for this work.  We gratefully acknowledge the support of the South Dakota Board of Regents Competitive Research Grant. We acknowledge support from the Russian Foundation for Basic Research, grant No.~15-02-02919. We acknowledge the support of the Natural Sciences and Engineering Research Council of Canada, funding reference number SAPIN-2017-00023, and from the Canada Foundation for Innovation John R.~Evans Leaders Fund.  This research used resources provided by the Oak Ridge Leadership Computing Facility at Oak Ridge National Laboratory and by the National Energy Research Scientific Computing Center, a U.S.~Department of Energy Office of Science User Facility. We thank our hosts and colleagues at the Sanford Underground Research Facility for their support.

\bibliography{reference.bib}

\end{document}